\documentclass[twocolumn,showpacs,superscriptaddress,preprintnumbers,amsmath,amssymb]{revtex4}
\usepackage{graphicx}
\usepackage{dcolumn}
\usepackage{bm}

\newcommand{\pt}{\ensuremath{p_{\mathrm{T}}}}
\newcommand{\tf}{\ensuremath{T_{\mathrm{f}}}}
\newcommand{\roots}{\ensuremath{\sqrt{s_{_{NN}}}}}
\newcommand{\mub}{\ensuremath{\mu_{B}}}
\newcommand{\tc}{\ensuremath{T_{\mathrm{c}}}}
\newcommand{\epsc}{\ensuremath{\epsilon_{\mathrm{c}}}}
\newcommand{\agev}{\ensuremath{A~\mathrm{GeV}}}
\newcommand{\gev}{\ensuremath{\mathrm{GeV}}}

\newcommand{\npart}{\ensuremath{\langle N_{\mathrm{part}} \rangle}}
\begin{document}

\preprint{}

\title{Hadronization and Hadronic Freeze-Out in Relativistic Nuclear Collisions}

\author{Francesco Becattini}
\affiliation{Universit\`a di Firenze and INFN Sezione di Firenze, Italy}

\author{Marcus Bleicher}
\affiliation{Frankfurt Institute for Advanced Studies (FIAS), Frankfurt, Germany}

\author{Thorsten Kollegger}
\affiliation{Frankfurt Institute for Advanced Studies (FIAS), Frankfurt, Germany}

\author{Michael Mitrovski}
\affiliation{Brookhaven National Laboratory, New York, USA}

\author{Tim Schuster\footnote{now at Yale University, New Haven, CT, USA.}}
\affiliation{Frankfurt Institute for Advanced Studies (FIAS), Frankfurt, Germany}
\affiliation{Institut f\"ur Kernphysik, J. W. Goethe-Universit\"at, Frankfurt, Germany}

\author{Reinhard Stock}
\affiliation{Frankfurt Institute for Advanced Studies (FIAS), Frankfurt, Germany}
\affiliation{Institut f\"ur Kernphysik, J. W. Goethe-Universit\"at, Frankfurt, Germany}

\date{\today}

\begin{abstract}
We analyze hadrochemical freeze-out in central Pb+Pb collisions at CERN SPS energies, employing the hybrid version of the Ultrarelativistic Quantum Molecular Dynamics model, which describes the transition from a hydrodynamic stage to hadrons by the Cooper-Frye mechanism, and matches to a final hadron-resonance cascade. We fit the results both before and after the cascade stage using the statistical model, to assess the effect of the cascade phase. We observe a strong effect on antibaryon yields except anti-$\Omega$, resulting in a shift in T and \mub\ of the freeze-out curve. We discuss indications of a similar effect in SPS and RHIC data, and propose a method to recover the bulk hadron freeze-out conditions.
\end{abstract}

\pacs{25.75.-q,25.75.Nq,24.85.+p,24.10.Pa,24.10.Nz}

\maketitle

\section{Introduction}

Hadron production in relativistic A+A collisions is supposed, since Bevalac times~\cite{1}, to proceed via two separate stages. The first, ``hadrochemical'' freeze-out fixes the bulk hadronic yields per species which are conserved throughout the subsequent hadron-resonance cascade expansion. At its end, ``kinetic freeze-out'' delivers
the eventually observed bulk properties such as \pt\ spectra, HBT correlations,
collective flow, etc.. Most remarkably, the hadronic yield distributions over species are understood to resemble grand canonical statistical Gibbs equilibrium ensembles~\cite{2,3}, from AGS up to RHIC and LHC energies.
The two most relevant parameters, temperature $T$ and baryochemical potential \mub\ thus capture a snapshot of the system dynamical evolution, taken at the instant of hadrochemical freeze-out.

In relativistic A+A collisions the thus determined $T$ increases monotonically with \roots, saturating at about 170~MeV while \mub\ approaches zero. Systematic statistical model (SM) analysis reveals the ``freeze-out curve''~\cite{4} in the $T,\mub$ plane, in which we usually also represent the conjectured plot of the phase diagram of QCD matter.

\begin{figure}
\includegraphics[width=0.39 \textwidth]{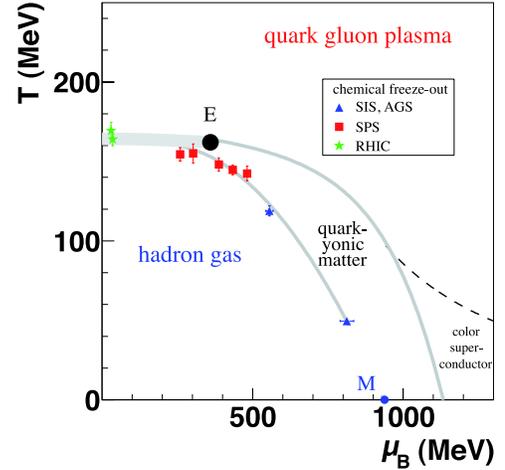}
\caption{(Color online) Sketch of the QCD phase diagram, including the hadronic freeze-out curve (see text).}
\label{fig:fig1}
\end{figure}

Such a plot is given in Fig.~\ref{fig:fig1}. It shows two principal lines, firstly a parton-hadron coexistence boundary line, inferred from lattice QCD~\cite{5} at low \mub, and from chiral restoration theory~\cite{6} at high \mub. And, second, the SM freeze-out curve. Remarkably, the lines merge toward $T=170$~MeV, $\mub=0$.
The freeze-out curve thus locates the QCD hadronization transition temperature \tc: hadronization 
seems to coincide with hadronic freeze-out, here~\cite{7}. Equally remarkable, however, the two lines disentangle with
increasing \mub, becoming spaced by about 30~MeV temperature difference
toward $\mub=500$~MeV which corresponds to $\roots=5$~GeV in A+A collisions.

What are we freezing out from, here? If hadronization coincides with the parton-hadron transition 
line at small \mub, and if hadronization involves a statistical equilibrium
of hadron species \cite{various,8}, one could imagine that the success of the statistical model in reproducing hadronic multiplicities in the higher \mub\ region 
indicates that the chemical freeze-out line marks the transition to the hadron world from 
an alternate phase. Indeed, the possible existence of an alternative QCD phase, so-called quarkyonic matter~\cite{9}
has been put forward based on similar arguments \cite{allofthem}. Indicated in Fig.~\ref{fig:fig1} 
is a scenario in which the freeze-out curve is 
identified, tentatively, with a hypothetical quarkyonic matter phase boundary.

Before embarking on this idea a different possible situation needs to be addressed. Taking for granted that the hadron-resonance phase is indeed created directly at the coexistence curve it might be conceivable that an expansive hadron/resonance evolution stage, setting in at hadronization temperature and baryochemical potential \tc\ and $\mu_{B,\mathrm{c}}$, cools down the population maintaining chemical equilibrium until chemical freeze-out occurs by mere dilution (the inelastic mean free path becoming larger than the system size), at lower $T$ and higher \mub, thus defining the freeze-out curve.

In this paper we test the latter scenario. We employ the framework of the Ultrarelativistic Quantum Molecular Dynamics (UrQMD) microscopic transport model. Its recent hybrid version~\cite{10} features a 3+1-dimensional hydrodynamic expansion during the high density stage, terminated by the Cooper-Frye mechanism once the energy density of flow cells falls below a ``critical'' energy density, assumed to be $\epsc=0.8$~GeV/fm$^3$. This criterion resembles the parton-hadron transition line of Fig.~\ref{fig:fig1}.
The hadron/resonance population can be examined, either, by terminating the model evolution at this stage, emitting into vacuum, and fitting the yield
distribution by the grand canonical statistical model~\cite{11}. Or, alternatively, the UrQMD hadron/resonance cascade expansion stage is attached, as an ``afterburner''. The outcome is again fitted by the SM.
A preliminary report on this work was published in Ref.~\cite{becattini:qm2011}.
We shall show that the afterburner does not cool the system in equilibrium (to start from \tc\ and arrive at a new equilibrium \tf\ on the freeze-out curve).
However, it is shown to selectively absorb antiprotons and antihyperons except anti-$\Omega$s at SPS energies, while leaving the other yields essentially unchanged. A distorted yield distribution results, as it was already shown by Bass and Dumitru~\cite{12}.
We shall explore a method to circumvent such final state effects, recovering the true hadronic freeze-out curve.

\section{UrQMD}

\begin{figure}
\includegraphics[width=0.39 \textwidth]{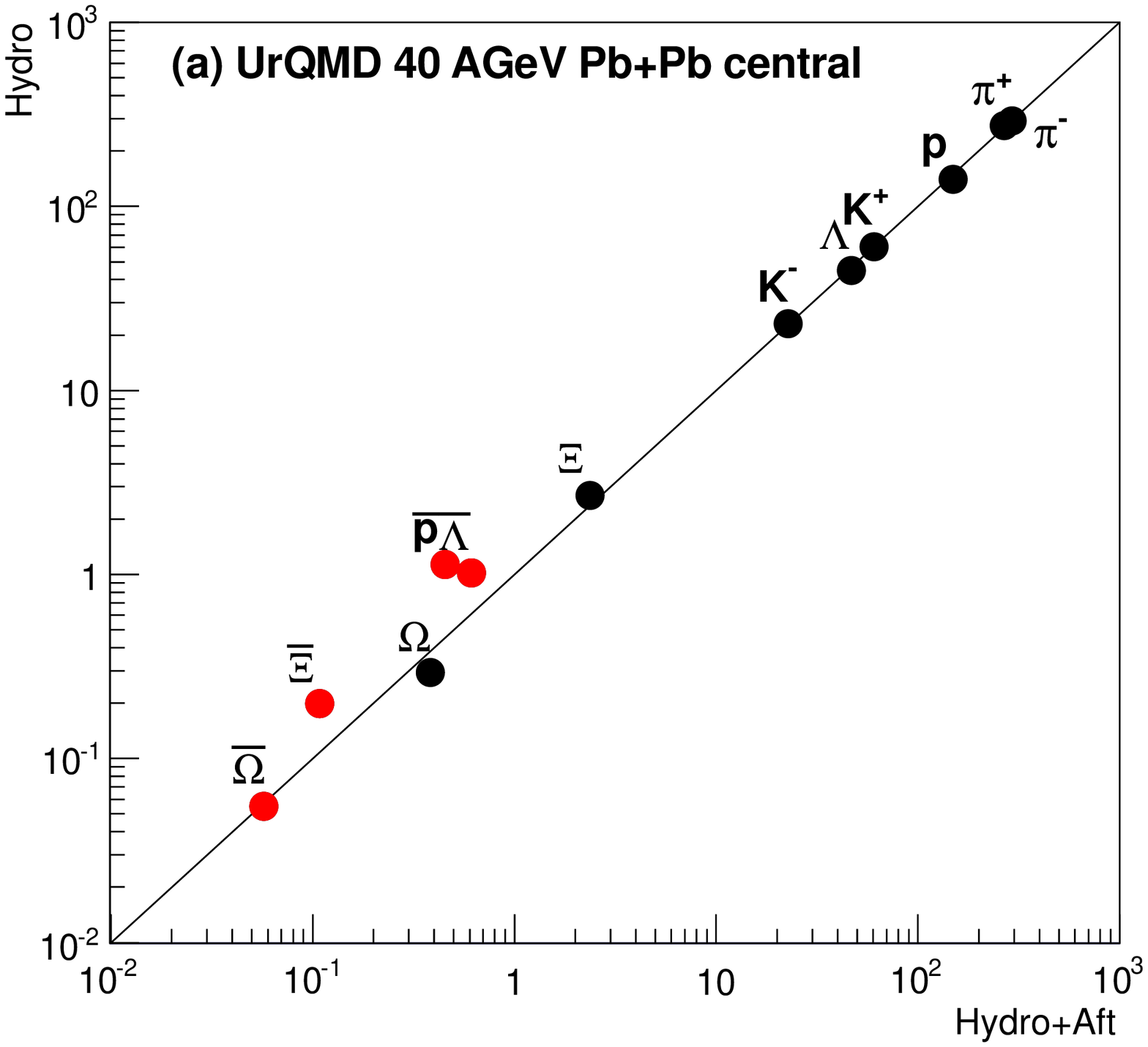}
\includegraphics[width=0.39 \textwidth]{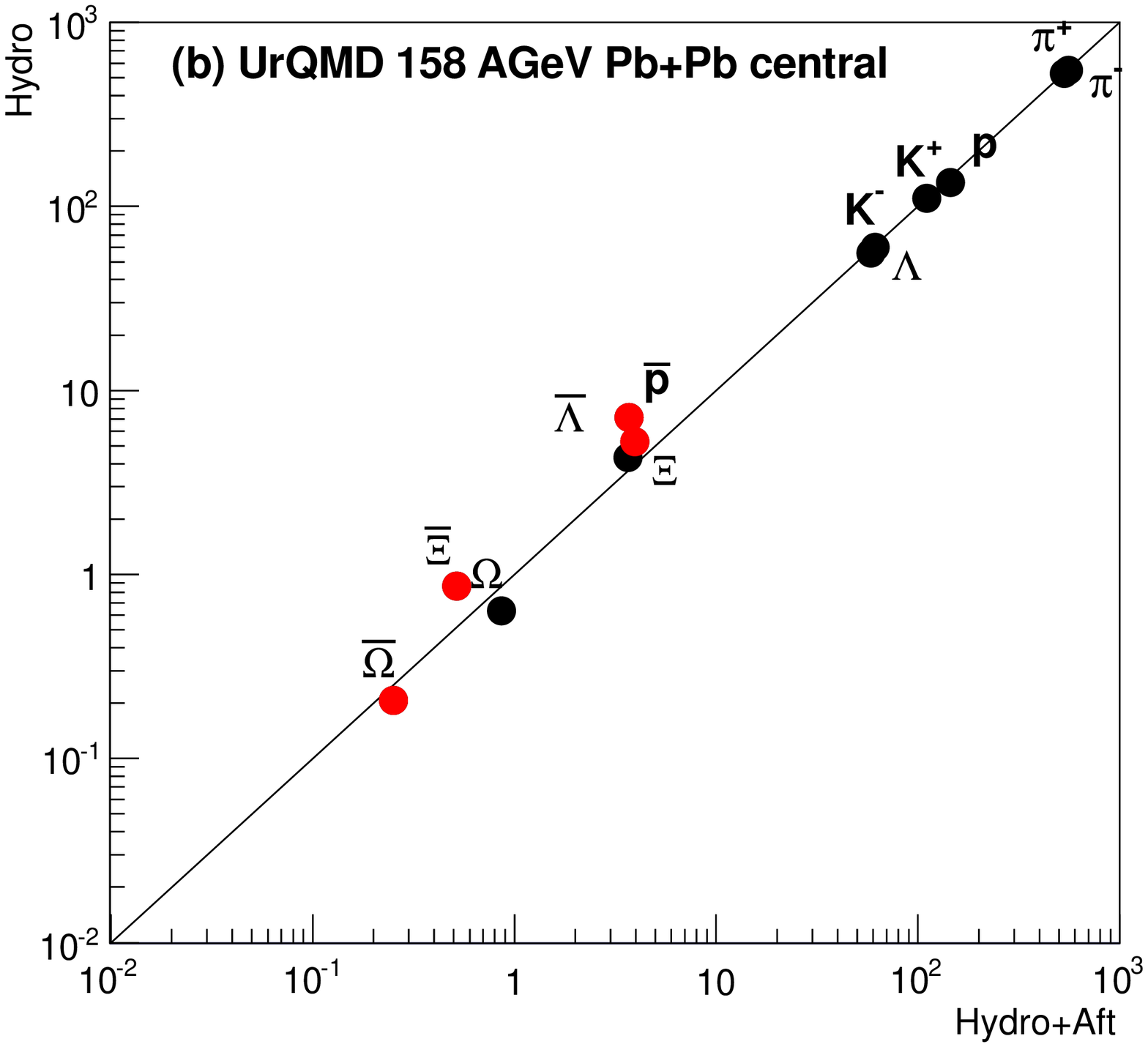}
\caption{(Color online) UrQMD calculations for hadron multiplicities in central Pb+Pb collisions at (a) 40$A$ and (b) 158$A$~GeV, plotting the result before the hadronic cascade vs.\ the result after. Antibaryons are highlighted by red markers.}
\label{fig:fig2}
\end{figure}

Figure~\ref{fig:fig2} shows the effect of the final UrQMD cascade stage, in a plot of hadron multiplicities directly after the hydro stage, vs.\ the multiplicities at the end of the cascade. We illustrate these conditions for the 5\% most central Pb+Pb collision multiplicities, at the SPS energies 40 and 158\agev\ ($\roots = 8.6$ and $17.3~\gev$). We see the bulk hadrons unaffected by the afterburner, including the $\Xi$ and anti-$\Omega$. Whereas the other antibaryons, $\bar{\mathrm{p}}$, $\bar{\Lambda}$ and $\bar{\Xi}$, are significantly, and selectively suppressed. We quantify this suppression in Fig.~3 showing the various hadron multiplicity ratios, of full UrQMD calculations (``Hydro+Afterburner''), relative to the output after Cooper-Frye transition (``Hydro''), for the two cases illustrated in Fig.~\ref{fig:fig2}.
The $\bar{\mathrm{p}}$, $\bar{\Lambda}$ and $\bar{\Xi}$ multiplicities show reductions ranging from 45 to 25\%, whereas the other hadron yields stay essentially constant. Similar results are reported in ref.~\cite{12} where it was shown, furthermore, that at top RHIC and LHC energies the selective absorption of these antibaryon species turns into a general suppression of both the corresponding baryon and antibaryon species. This should be a consequence of the approximate particle-antiparticle symmetry occurring at such high energies, where \mub\ approaches zero.

The mechanism of this suppression of antibaryons (at SPS energies), in the course of the UrQMD hadron/resonance cascade phase, appears to be the excitation to higher resonance states which increases the inelastic (annihilation) interaction rate, over and above ``direct'' antibaryon-baryon annihilation. The resonance mechanism should not be active for the $\Omega$ hyperons which lack such low lying excitations.

\begin{figure}
\includegraphics[width=0.39 \textwidth]{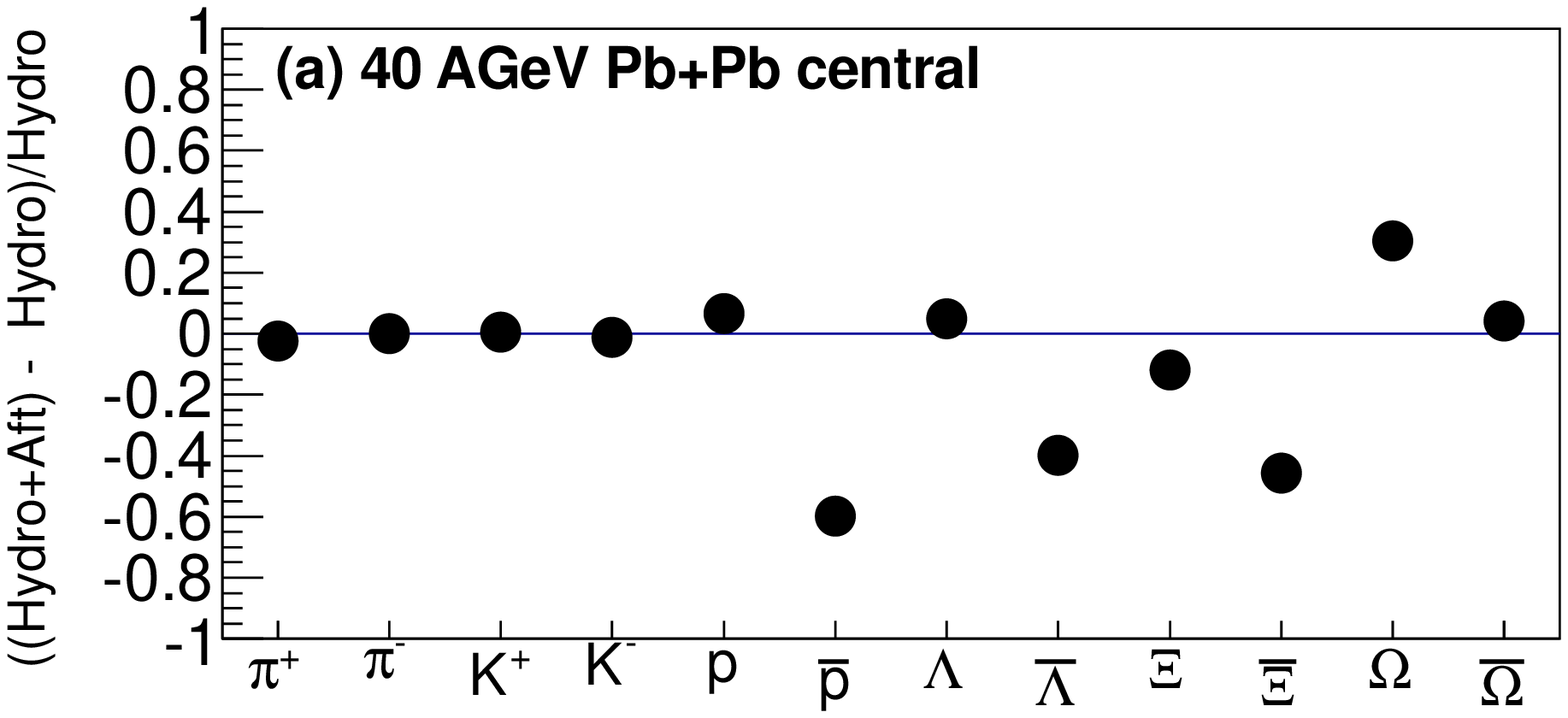}
\includegraphics[width=0.39 \textwidth]{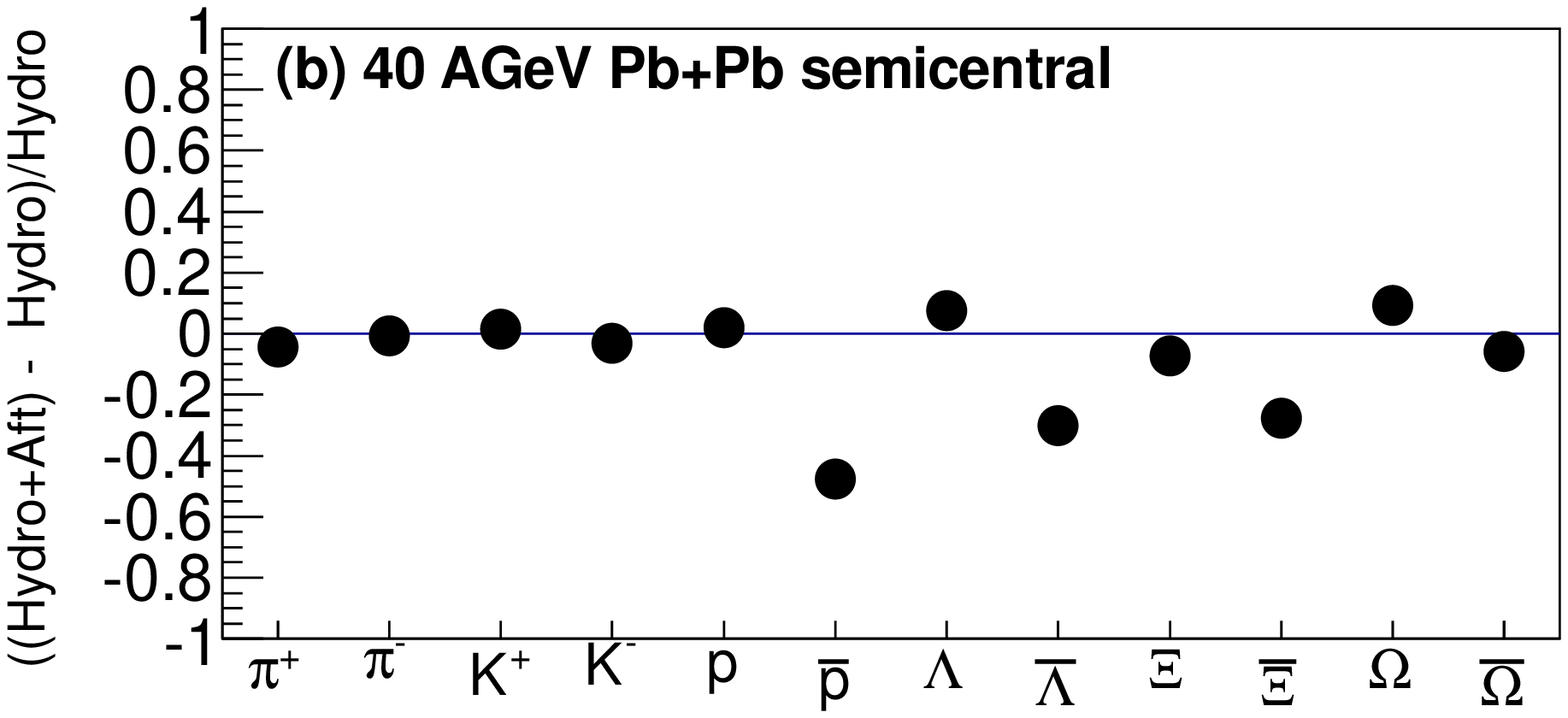}
\includegraphics[width=0.39 \textwidth]{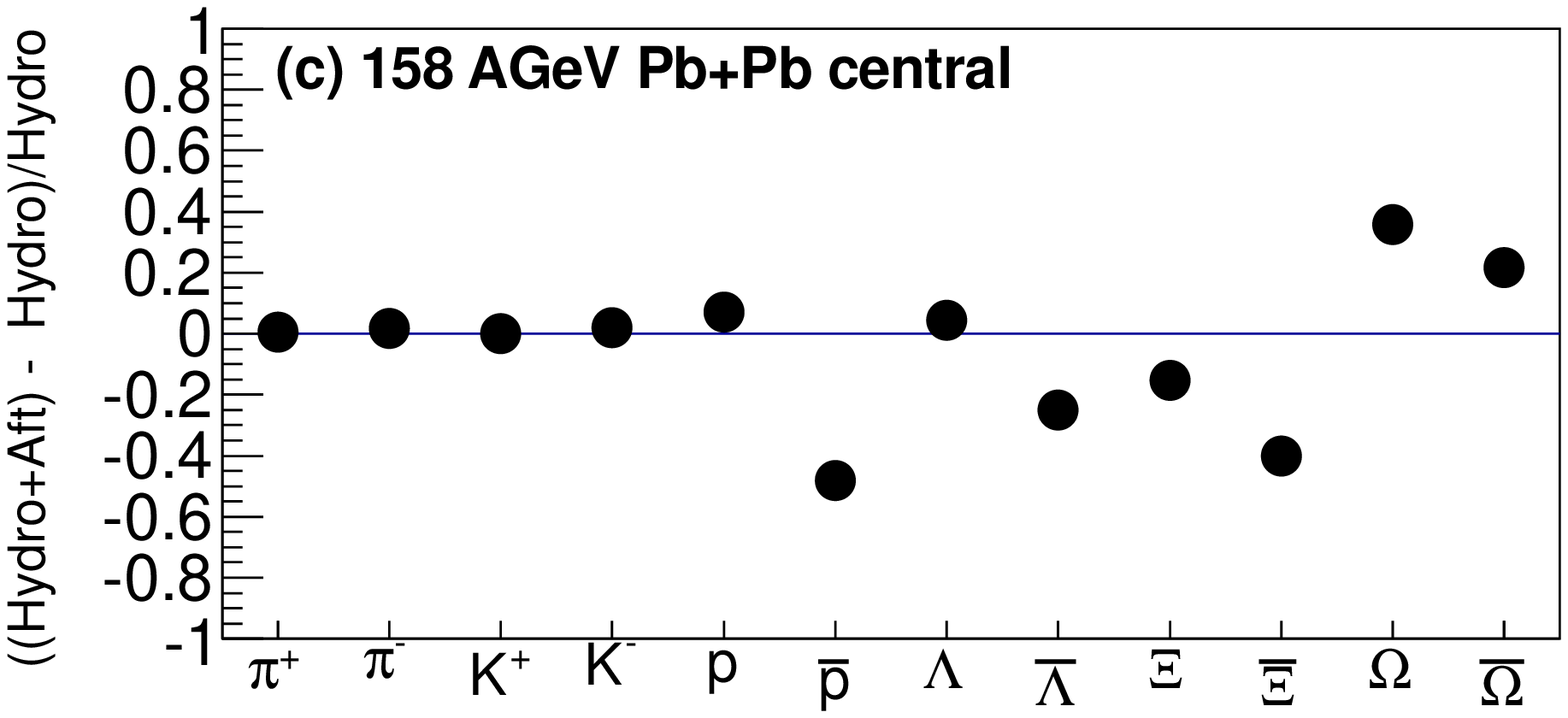}
\includegraphics[width=0.39 \textwidth]{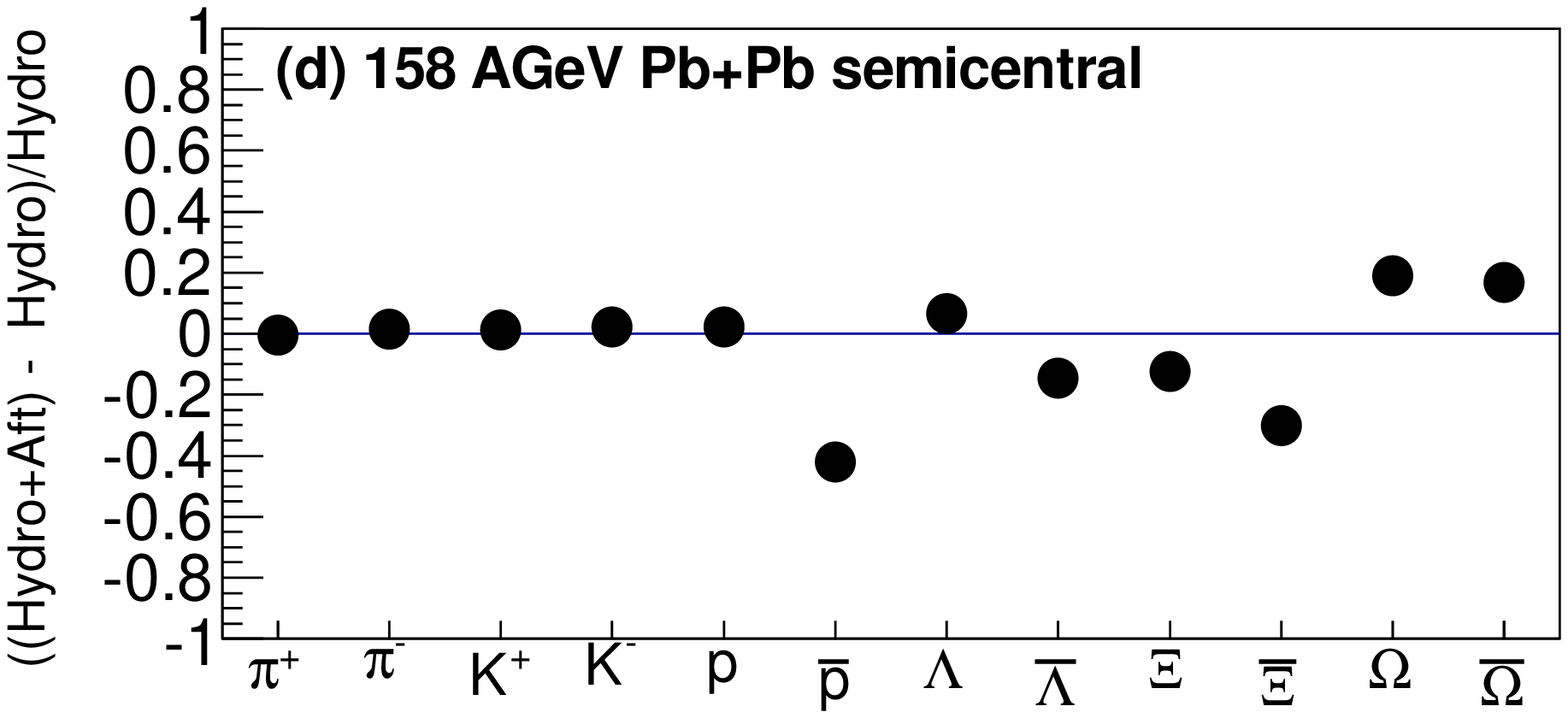}
\caption{The ratios of hadron multiplicities after the UrQMD cascade stage, to the
multiplicities directly after hadronization (after the hydro-stage). We illustrate central Pb+Pb collisions at (a) 40 and (c) 158\agev, as well as the corresponding semi-central cases for $\npart = 90$ (b), (d).}
\label{fig:fig3}
\end{figure}

Figure~\ref{fig:fig3} also illustrates similar UrQMD results obtained for semi-central Pb+Pb collisions in the 35 to 45 percent centrality bin where \npart\ is about 90. Qualitatively similar suppression occurs, albeit of reduced magnitude owing to the shorter duration of the final cascade stage. We shall return in Sec.~\ref{sect:data} to experimental consequences of this annihilation increase toward central collisions.

\section{Statistical Model Analysis and the Freeze-Out Curve}

Turning to statistical model analysis of the UrQMD results we consider the total hadron multiplicities in central Pb+Pb collisions at the SPS energies 20, 30, 40, 80 and 158\agev\ ($\roots = 6.3, 7.6, 8.7, 12.3$ and $17.3~\gev$).
They are chosen to obtain a sufficient coverage of the freeze-out curve
illustrated in Fig.~\ref{fig:fig1}, and to resemble the positions of the data points
gathered by NA49~\cite{13,17}. In order to obtain results under statistical conditions
similar to those prevailing in the analysis of the corresponding data we attach
to the hadron multiplicity results from UrQMD  the corresponding total error resulting from the addition in quadrature of the statistical and systematic errors reported by NA49~\cite{13,17}.
We employ the grand canonical 
version of the statistical model as described in refs.~\cite{2,11}. 
In this version, the statistical model is supplemented with a free parameter $\gamma_S$ suppressing 
the production of hadrons containing $n_s$ valence strange quarks according to $\gamma_S^{n_s}$.
In the SPS energy region, as well as in the peripheral collisions at RHIC energy \cite{star,becarhic}
this parameter turns out to be less than one~\cite{2,11}. It has been observed that this parameter 
can be explained, at least in the RHIC energy region, with the superposition of two sources (the core-corona model): single nucleon-nucleon (NN) collisions and a fully equilibrated source~\cite{becacc}.
For all hydro+UrQMD calculations, this parameter is set to 1 and no core-corona effect is 
included.

\begin{figure}
\includegraphics[width=0.39 \textwidth]{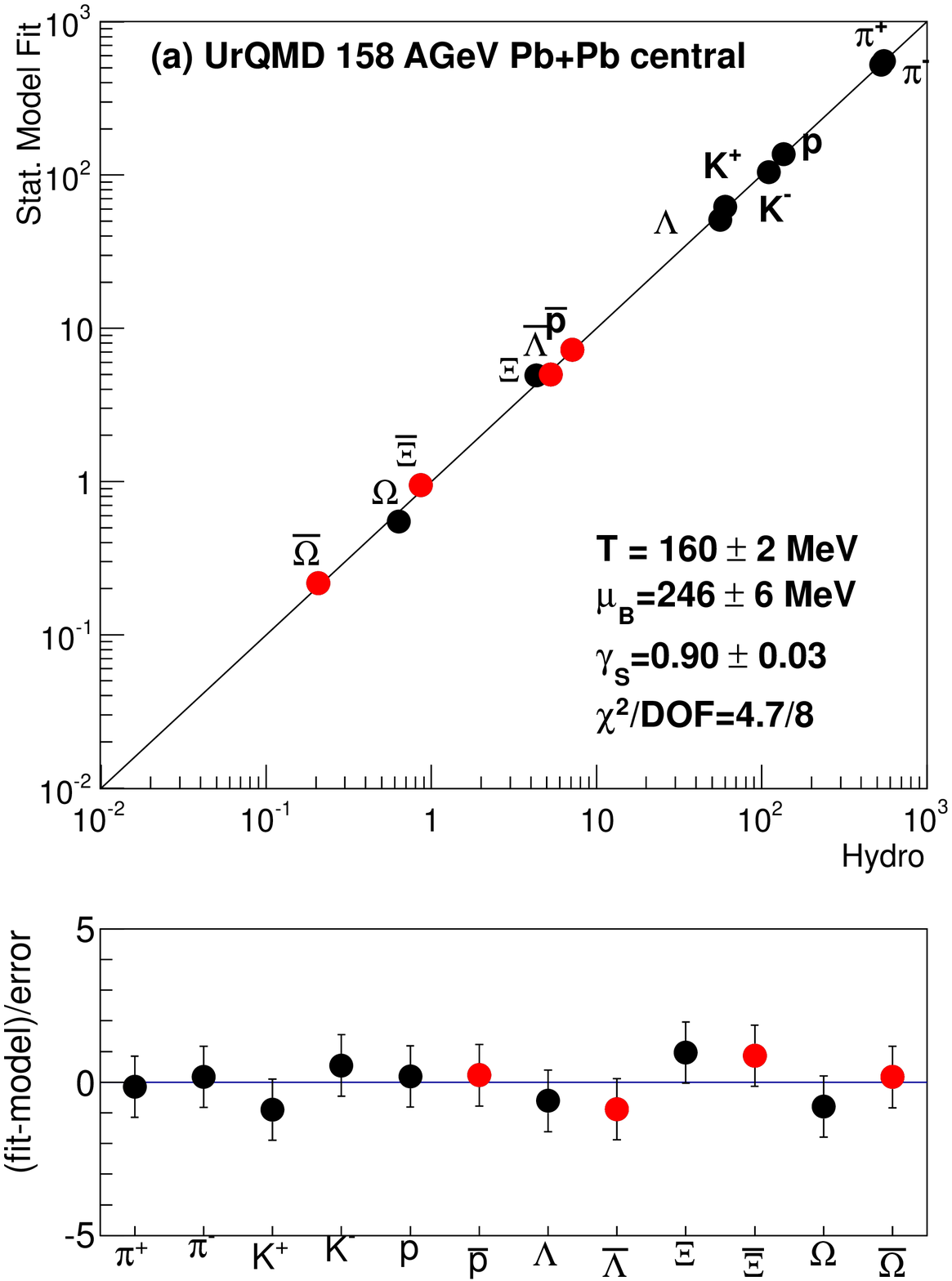}
\includegraphics[width=0.39 \textwidth]{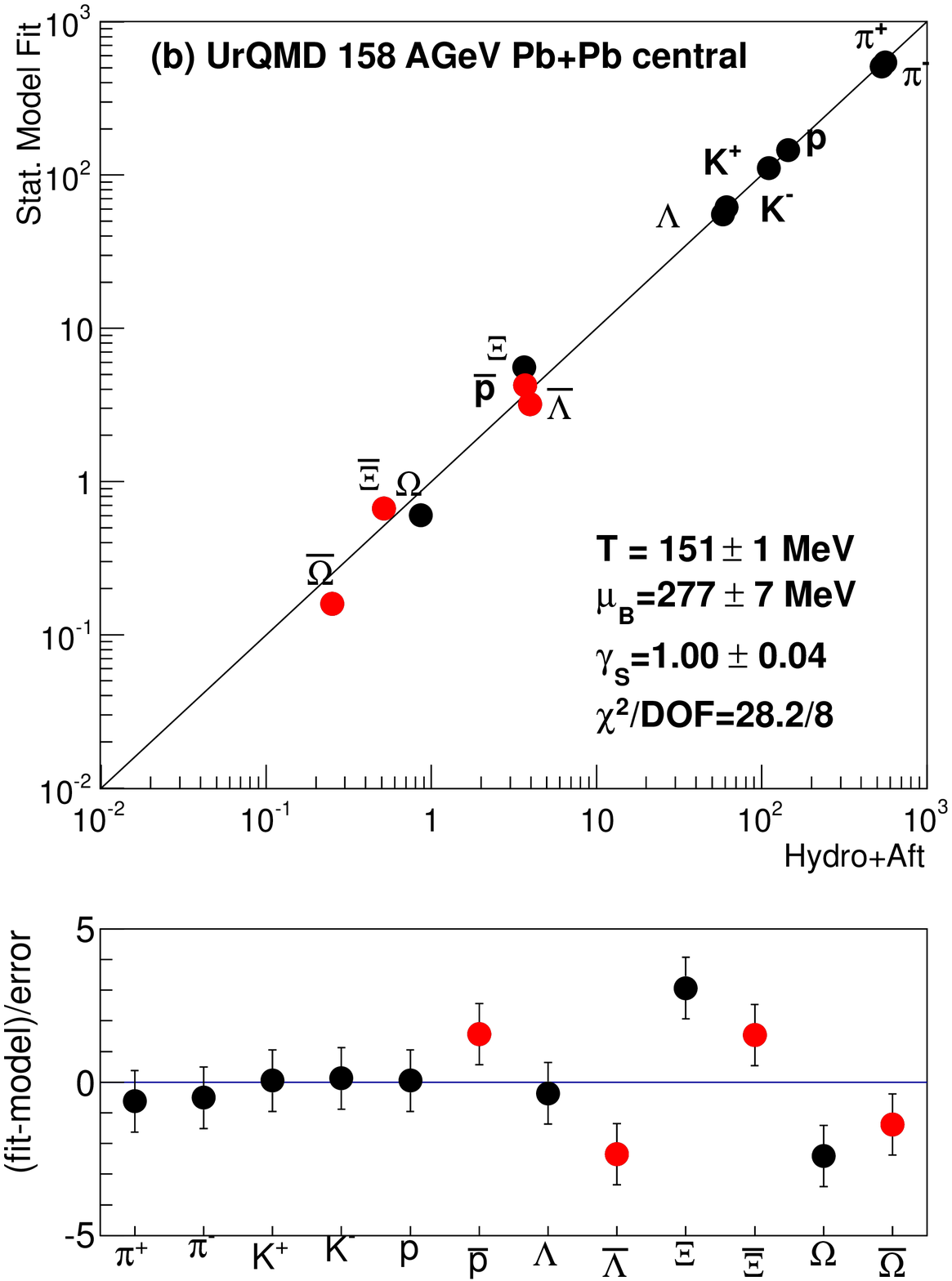}
\caption{(Color online) Statistical model fit to UrQMD results for hadron multiplicities in Pb+Pb central collisions at 158\agev. (a) employs UrQMD terminated directly after the hydro stage whereas (b) refers to multiplicities obtained after the final hadron/resonance cascade phase of UrQMD.}
\label{fig:fig4}
\end{figure}

\begin{figure}
\includegraphics[width=0.39 \textwidth]{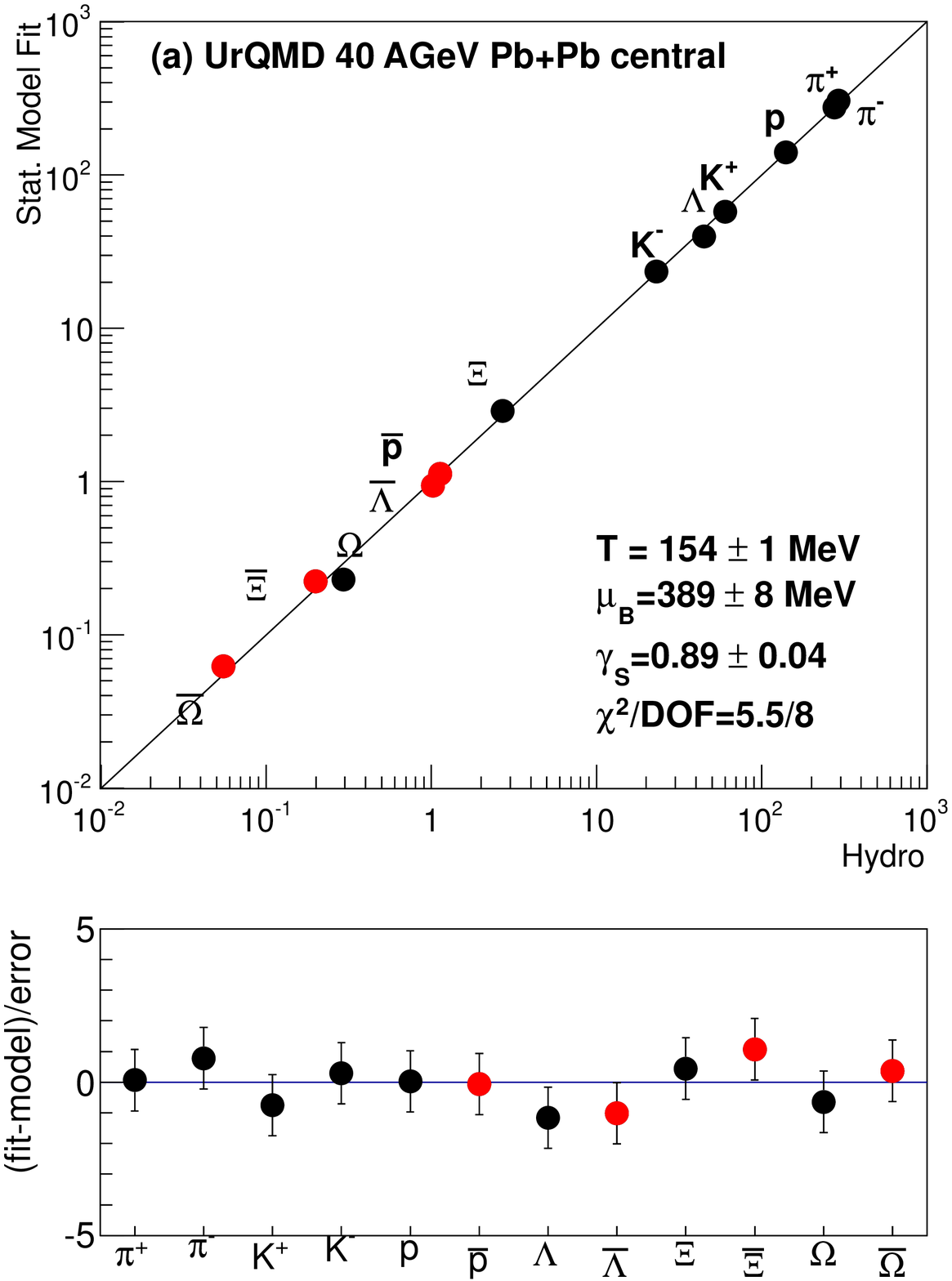}
\includegraphics[width=0.39 \textwidth]{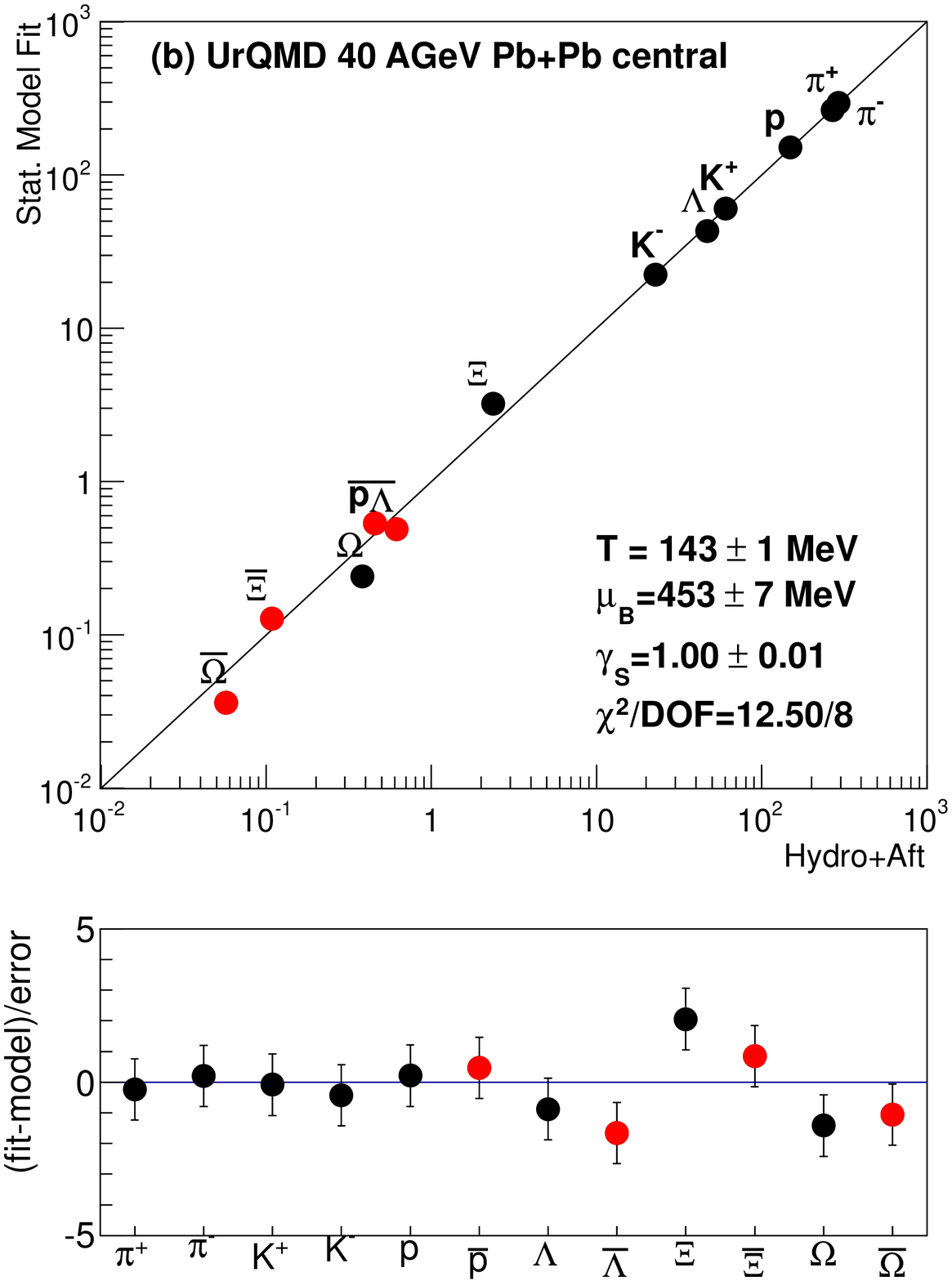}
\caption{(Color online) Same as Fig.~\ref{fig:fig4} but for Pb+Pb collisions at 40\agev.}
\label{fig:fig5}
\end{figure}

Figure~\ref{fig:fig4} illustrates the fits to the hybrid UrQMD results by the statistical model, choosing the 158\agev\ case as an example. The afterburner stage indeed shifts $(T, \mub)$ considerably, from (160~MeV, 246~MeV) to (151~MeV, 277~MeV).
However, note the dramatic decrease of fit quality, from 4.7 to 28.2
in chisquare. Similarly, the results for the energy 40\agev\ are shown in Fig~\ref{fig:fig5}, exhibiting the same shift conditions. The effect of the afterburner is, thus, not an in-equilibrium cooling but rather a distortion of the hadron yield distribution in the antibaryon sector, away from equilibrium - as we could guess from Figs.~\ref{fig:fig2} and~\ref{fig:fig3}, already.
The parameter $\gamma_S$ is very close to 1 throughout.

\begin{figure}[h!]
\includegraphics[width=0.39 \textwidth]{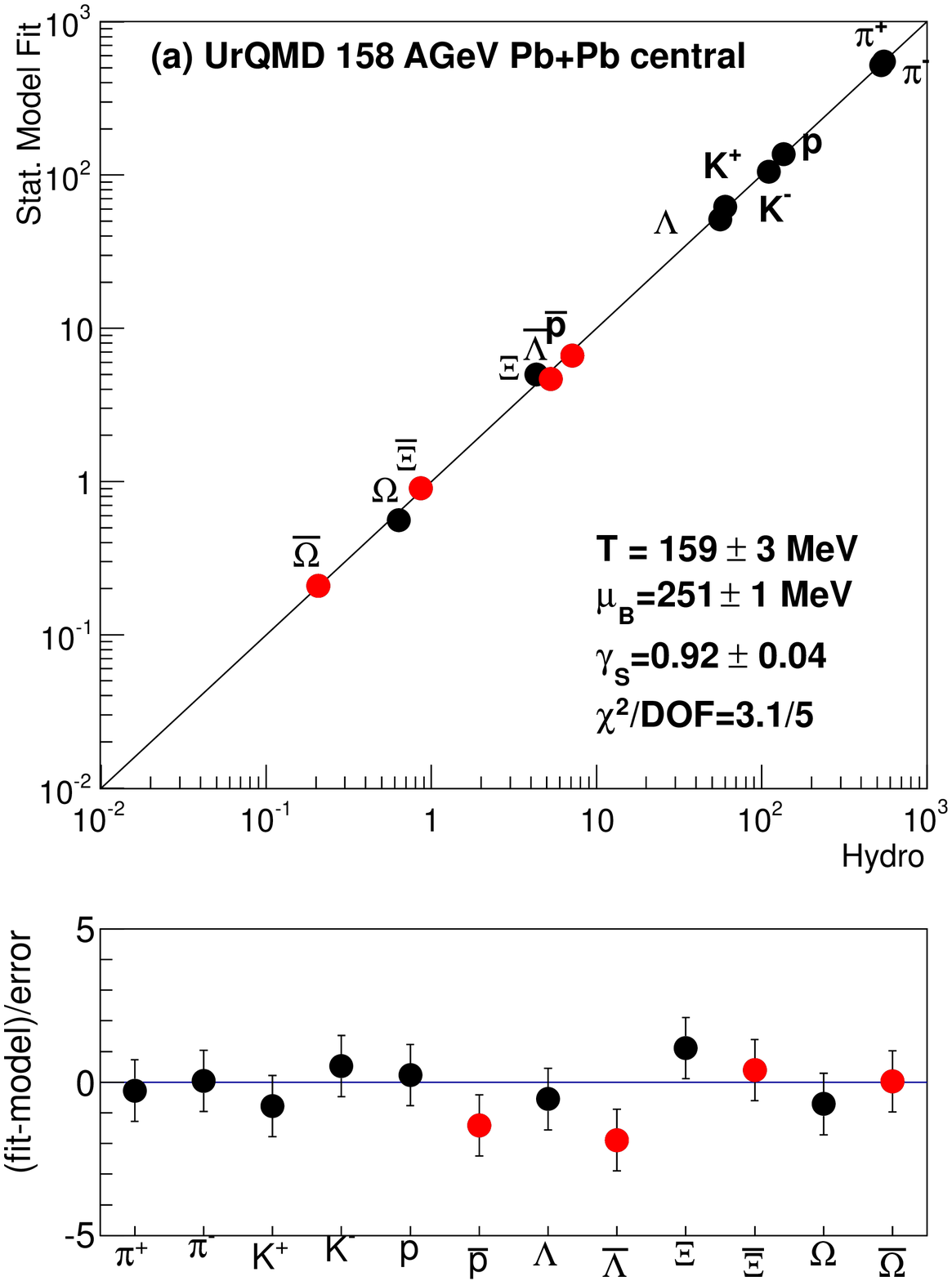}
\includegraphics[width=0.39 \textwidth]{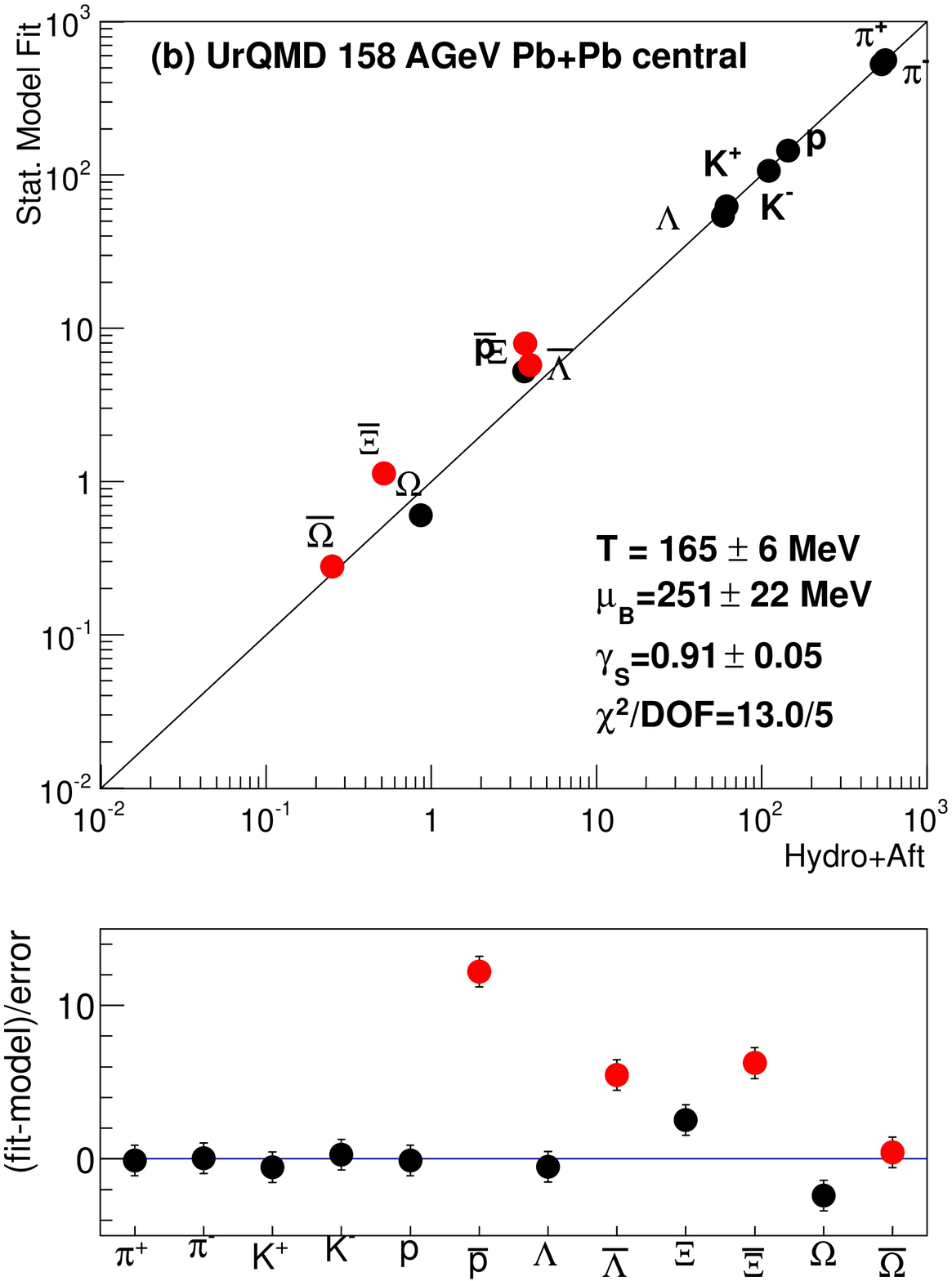}
\caption{(Color online) Same as in Fig.~\ref{fig:fig4} but excluding $\bar{\mathrm{p}}$, $\bar{\Lambda}$ and $\bar{\Xi}$ from the statistical model fit. Note that the residual plots do nevertheless also illustrate the entries resulting for these antibaryons.}
\label{fig:fig6}
\end{figure}

\begin{figure}[h!]
\includegraphics[width=0.39 \textwidth]{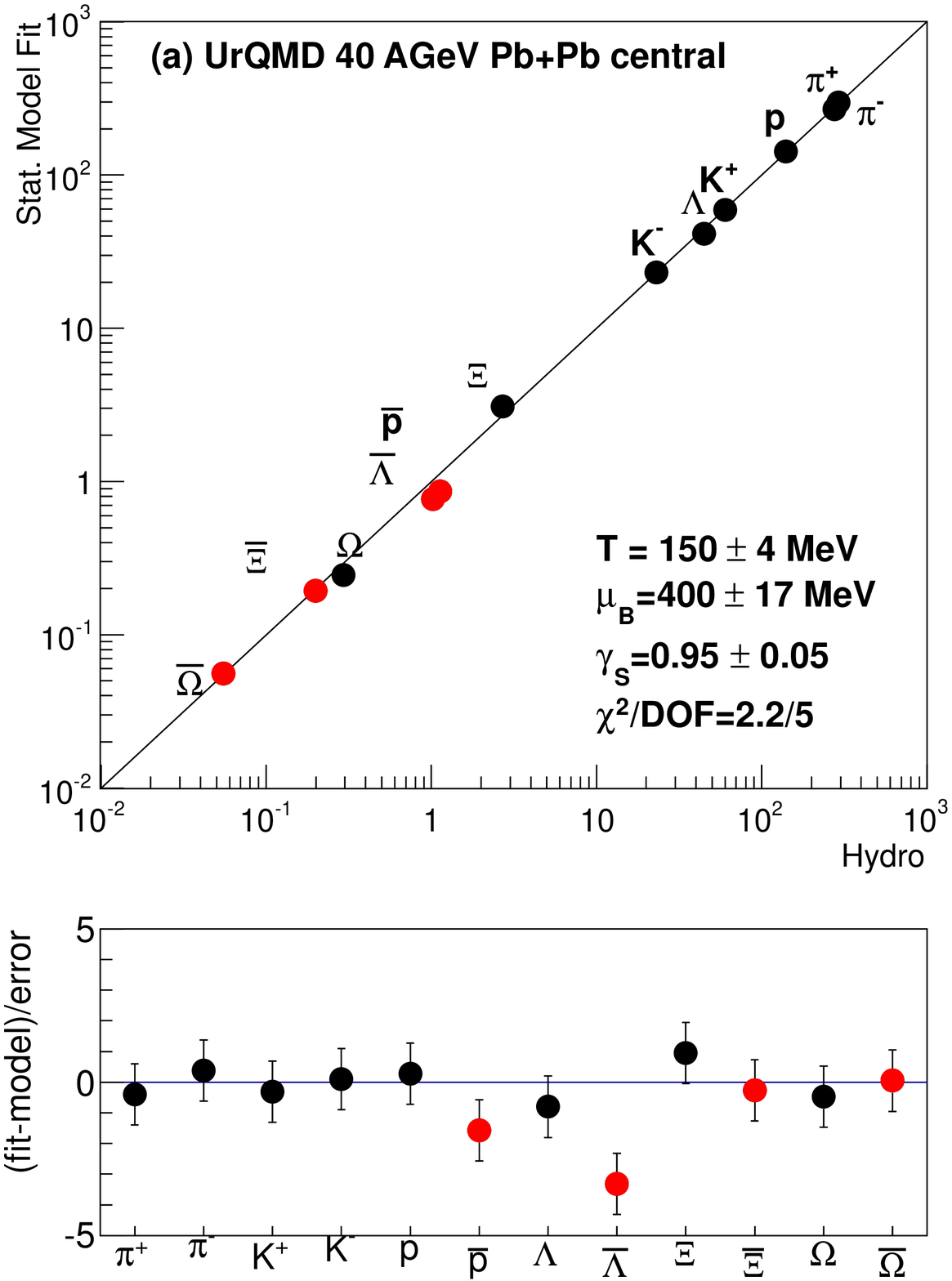} 
\includegraphics[width=0.39 \textwidth]{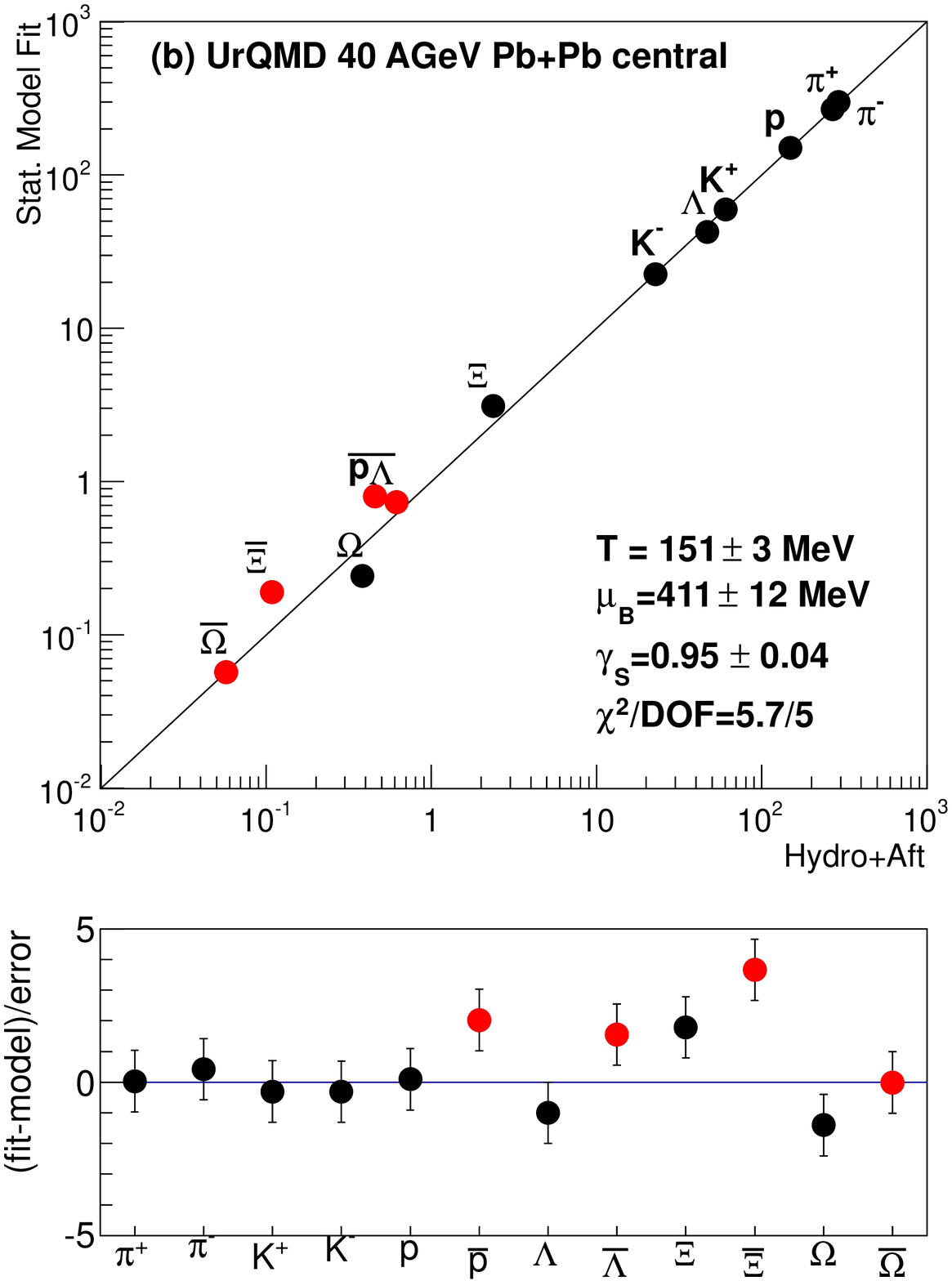}
\caption{(Color online) Same as Fig.~\ref{fig:fig6} for Pb+Pb at 40\agev.}
\label{fig:fig7}
\end{figure}

\begin{figure}
\includegraphics[width=0.39 \textwidth]{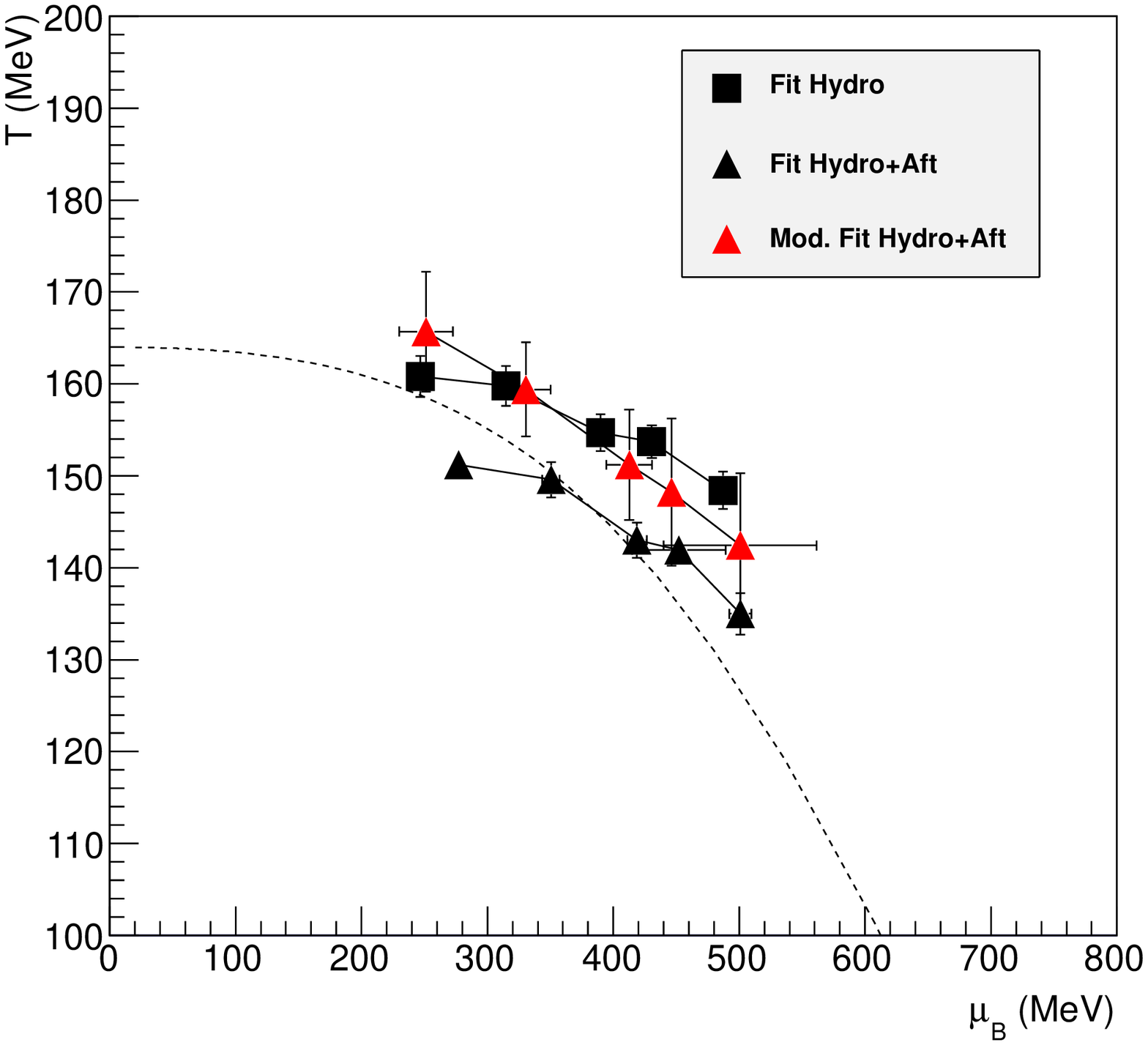}
\caption{(Color online) Summary of the results obtained for central Pb+Pb at five energies in the SPS range ($\roots = 6.3$, 7.6, 8.7, 12.3 and 17.3~GeV). $T$ and \mub\ are shown from statistical model fits, at each energy, to the three UrQMD configurations illustrated in Figs.~\ref{fig:fig4} to~\ref{fig:fig7}. The dashed line indicates the ``empirical freeze-out curve''~\cite{4}.
The fit to hadron multiplicities right after Cooper-Frye transition from the hydro stage (squares) results in higher temperatures. The afterburner stage of hadron/resonance rescattering cools the system (black triangles). The original freeze-out conditions are essentially recovered by the modified fit (red triangles) as described in the text.}
\label{fig:fig8}
\end{figure}

The idea arises to exclude $\bar{\mathrm{p}}$, $\bar{\Lambda}$ and $\bar{\Xi}$ from the SM fit. Figsures~\ref{fig:fig6} and~\ref{fig:fig7} show examples, again at 158 and 40\agev.
No cooling occurs. The fit to the afterburner output (which features a reasonable chisquare) now ignores the far off-diagonal $\bar{\mathrm{p}}$, $\bar{\Lambda}$ and $\bar{\Xi}$ entries.
A summary of the results, obtained for all energies, is given in Fig.~\ref{fig:fig8}. The $(T, \mub)$ positions of all analyzed cases are represented in a phase diagram that also exhibits the average freeze-out curve resulting from previous statistical model analysis of experimental data~\cite{4}. We see that the series of entries resulting from the full UrQMD calculation, including the final hadron/resonance cascade stage, follows rather closely the freeze-out curve. The points obtained without the cascade stage are generally well above, and also shifted to lower \mub. Not surprisingly, they approach the lattice parton-hadron boundary shown in Fig.~\ref{fig:fig1}. Most significantly, however, the points from full UrQMD but with omission of the $\bar{\mathrm{p}}$, $\bar{\Lambda}$ and $\bar{\Xi}$ multiplicities are also following the latter behavior: they universally exhibit no significant ``cooling'', with the exception, perhaps, of the high \mub\ region.

We conclude that the hadron/resonance cascade as modeled in the microscopic dynamics of UrQMD can not transport an initially established hadrochemical equilibrium from the phase coexistence line of Fig.~\ref{fig:fig1}, downward to the freeze-out line. However, it distorts the hadron yield distribution which leads to a downward shift of the freeze-out parameters derived from statistical model analysis, albeit at the cost of a rather unsatisfactory fit quality.
Far better fits are obtained omitting the three antibaryon species, which shifts the freeze-out curve upward.
We note, however, that it will be of key interest to extend the present study toward smaller \mub. Will the three curves in Fig.~\ref{fig:fig8} converge, more closely, toward $\mub=0$? This is a subject for ongoing study. However, both former~\cite{12} and concurrent~\cite{steinheimer} investigations with a hybrid UrQMD model conclude on significant afterburner effects, even up to LHC energies. Along with forthcoming data (see next section) the method proposed here may thus lead to a certain, general upward shift of the freeze-out curve at low \mub.

\section{A Look at the Data}
\label{sect:data}

We note that the properties reported above may be a consequence of the specific hadron transport model employed in the UrQMD approach, which implements a cascade of binary hadron/resonance collisions, combined with string excitation in such collisions which leads, in part, to multiparticle decays. However, the detailed-balance counterparts of the latter processes are not included. It has been argued repeatedly~\cite{8,14} that such multi-hadron collisions could significantly decrease the effective chemical relaxation constants governing antihyperon densities and their approach to chemical equilibrium. However, in these approaches only the direct vicinity of the critical temperature, $\tc= 160-170$~MeV,
has been addressed. Thus this line of argument does, in fact, refer to
the parton-hadron phase transformation itself. It attempts 
to recast the hypothesis of equilibrium establishment via the quantum mechanical mechanism of the phase transition~\cite{7,various,15} in terms of a quasi-classical microscopic dynamics of a mixed phase, consisting of Hagedorn resonances or string excitation modes and hadrons plus hadronic resonances, in some sense a quasi-classical microscopic
mixed phase, and phase transition model. Such models, however, work in the immediate vicinity of \tc\ only, i.e.\ in application to RHIC and LHC energies. In the present UrQMD study we address, specifically, the possibility of a hadrochemical equilibrium freeze-out at significantly lower temperature, and high baryochemical potential, where the freeze-out curve disentangles from the parton-hadron phase boundary.
It is our observation that the UrQMD model dynamics can not transport the system, from initially imprinted equilibrium by the Cooper-Frye mechanism, while maintaining chemical equilibrium adjusted dynamically to the rapidly falling temperature during expansion. This is in line with a former study~\cite{belkacem} of UrQMD equilibrium features in the same energy domain. Putting the system into a box, with constant energy density, it was shown that chemical equilibration requires upward of $25~\mathrm{fm}/c$, even under constant energy density conditions, not available during cascade expansion. From that study we would conclude that, conversely, an equilibrium initially established would not be substantially distorted during the afterburner phase.

\begin{figure}
\includegraphics[width=0.39 \textwidth]{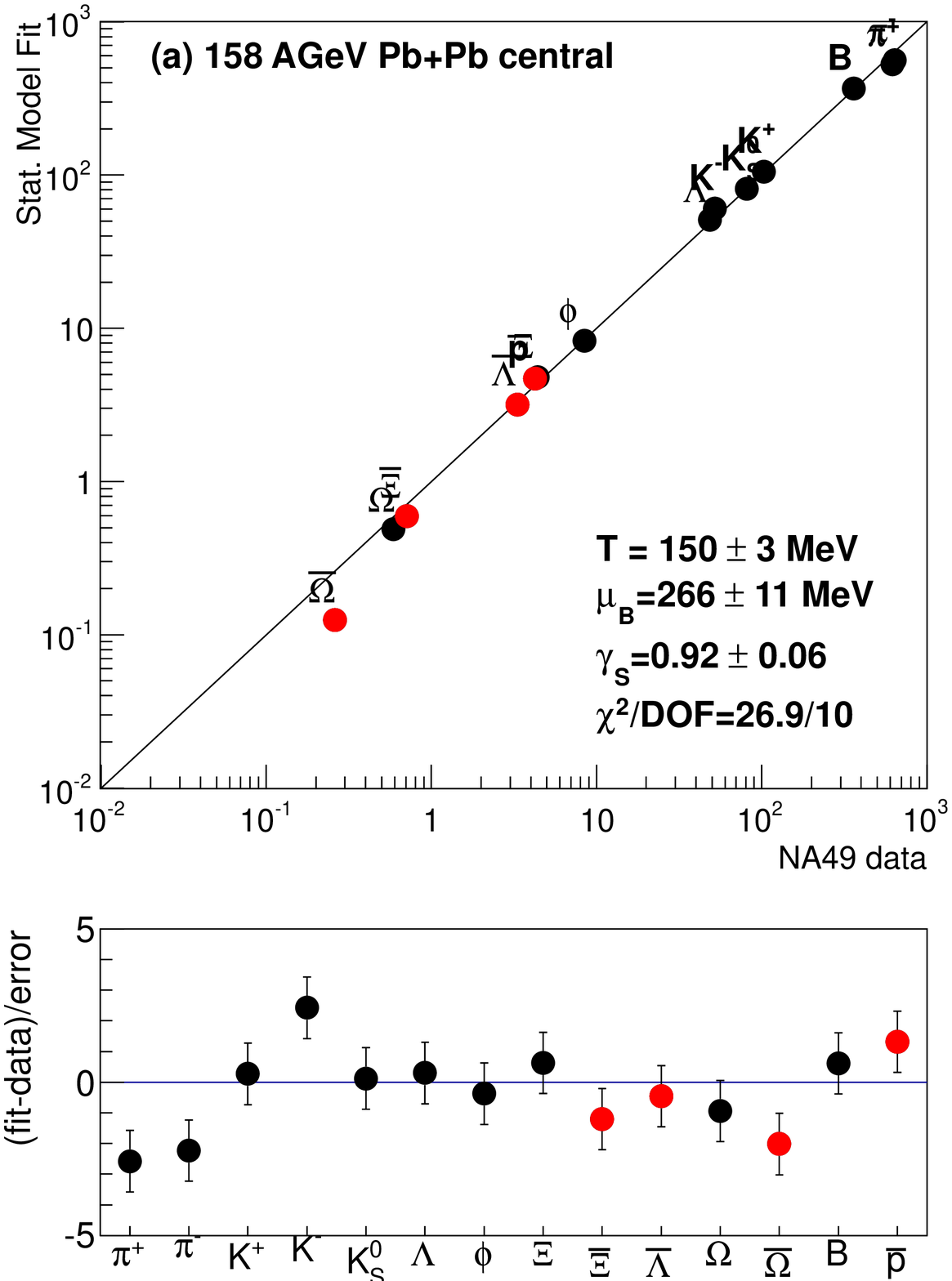}
\includegraphics[width=0.39 \textwidth]{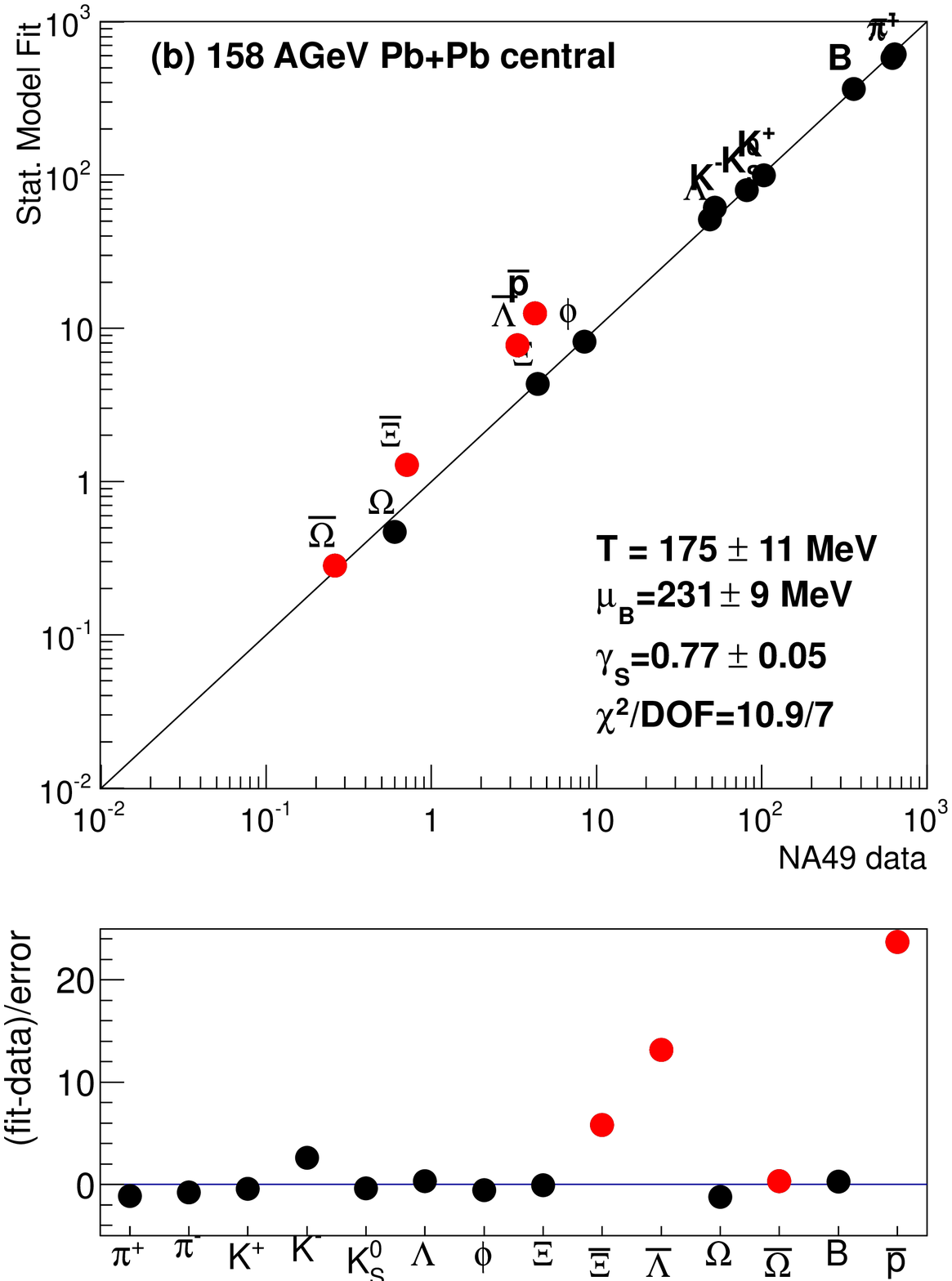}
\caption{(Color online) Statistical model fit to the NA49 data~\cite{13,17} for total hadron multiplicities in central Pb+Pb collisions at 158\agev. (a) full data set included in the fit. (b) fit with excluded $\bar{\mathrm{p}}$, $\bar{\Lambda}$ and $\bar{\Xi}$ entries (note that, nevertheless, the resulting residuals are illustrated).}
\label{fig:fig9}
\end{figure}

\begin{table}
\begin{center}
\caption{Main fit to hadron multiplicities (in full phase space) measured by NA49
 in central Pb+Pb collisions at $\sqrt s_{NN} = 17.3$~GeV~\cite{13,17} (Fig~\ref{fig:fig9}~(a)). Errors
 within brackets are the (realistic) parameter errors, obtained by rescaling the 
 original fit errors by $\sqrt{\chi^2/\mathrm{DOF}}$ according to the method suggested by the PDG~\cite{pdg}.}
\label{na49fit}

\begin{tabular}{ll}
   \multicolumn{2}{l}{Fit Parameters}  \\
   \hline
   \hline
   $T$ (MeV)       & 150.5$\pm$1.7 (3.1)       \\
   $\gamma_S$      & 0.915$\pm$0.034 (0.063)   \\
   $\mu_B$ (MeV)   & 266$\pm$6 (11)            \\
   $\chi^2/$DOF    & 26.9/10                    \\  
   \\
\end{tabular}

\begin{tabular}{lll}  
   Hadron        & Measured        & Fitted       \\ 
   \hline
   \hline
   $B$          &  362$\pm$8      &  367          \\
   $\pi^+$      &  619$\pm$35     &  528          \\ 
   $\pi^-$      &  639$\pm$35     &  560          \\  
   K$^+$        &  103$\pm$7      &  105          \\
   K$^-$        &  51.9$\pm$3.6   &  60.5         \\
   K$^0_S$      &  81$\pm$4       &  82           \\
   $\Lambda$    &  48.5$\pm$8.6   &  51.1         \\
   $\phi$       &  8.46$\pm$0.50  &  8.27         \\
   $\Xi^-$      &  4.40$\pm$0.64  &  4.80         \\
   $\bar{\rm p}$&  4.23$\pm$0.35  &  4.69         \\
   $\bar\Lambda$&  3.32$\pm$0.34  &  3.17         \\
   $\bar\Xi^+$  &  0.710$\pm$0.098&  0.592        \\
   $\Omega$     &  0.59$\pm$0.11  &  0.49         \\
   $\bar\Omega$ &  0.26$\pm$0.07  &  0.13         \\
\end{tabular}
\end{center}
\end{table}

In the $(T,\mub)$ domain treated here, the transport model employed in UrQMD may appear to be realistic. The idea thus arises to look for effects, similar to the findings summarized in Fig.~\ref{fig:fig8}, in real data. We recall from previous statistical model analysis of the SPS total multiplicity data~\cite{2,11}
that the obtained chisquare values tended to be rather high. We repeat here the SM analysis of the NA49 up-to-date 
data set (see Table~\ref{na49fit}) for central Pb+Pb collisions at top SPS energy, 158\agev\ ($\roots=17.3~\gev$).
With the updated data sample by NA49, the above trend is confirmed: the fit quality is worse than in
previous analyses as seen in Fig.~\ref{fig:fig9}~(a) and Table~\ref{na49fit}. This might be an indication
that the inelastic rescattering stage must be taken into account if one aims at matching the accuracy 
of the experimental measurements.
We have not repeated the analysis at lower SPS energies because the coverage of the hyperon sector was 
incomplete in NA49 but we note that new high precision results at $\roots=7.7~\gev$ are forthcoming 
from STAR, gathered in the recent low energy runs at RHIC~\cite{16}.

\begin{figure}
\includegraphics[width=0.39 \textwidth]{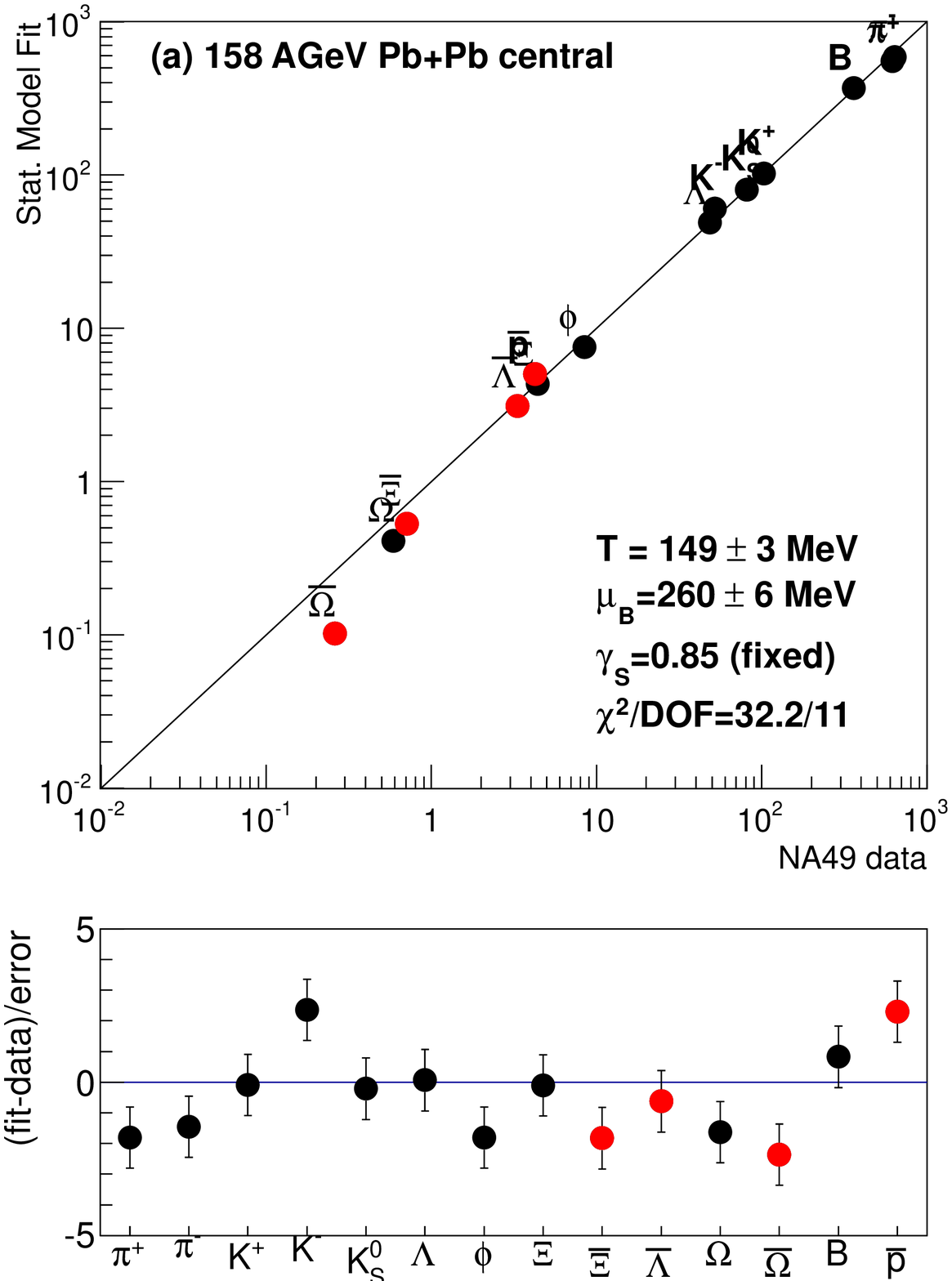}
\includegraphics[width=0.39 \textwidth]{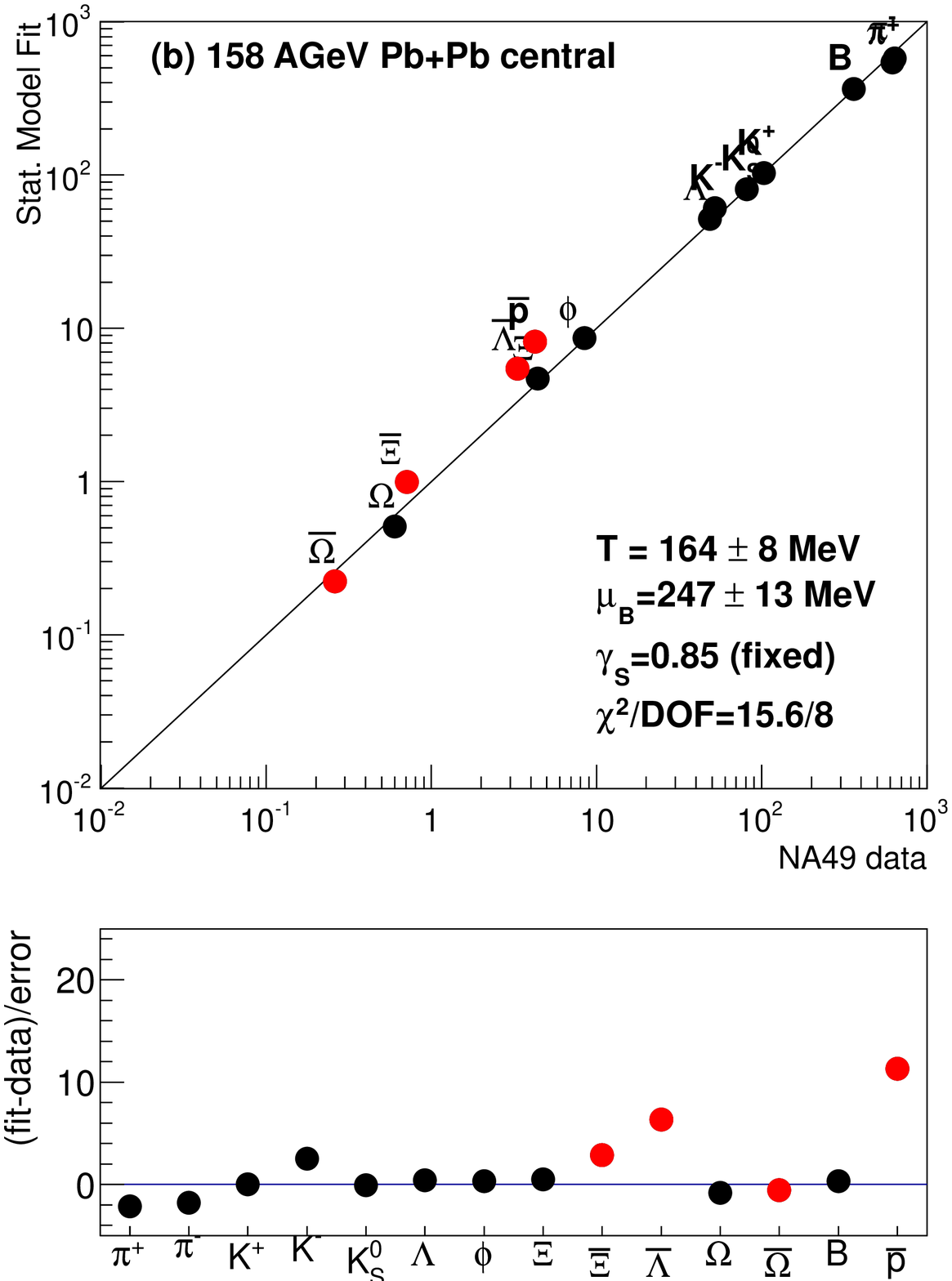}
\caption{(Color online) Statistical model fit to the NA49 data~\cite{13,17} for total hadron multiplicities in central Pb+Pb collisions at 158\agev with $\gamma_S= 0.85$ fixed. (a) full data set included in the fit. (b) fit with excluded $\bar{\mathrm{p}}$, $\bar{\Lambda}$ and $\bar{\Xi}$ entries (note that, nevertheless, the resulting residuals are illustrated).}
\label{fig:fig10}
\end{figure}

We show in Fig.~\ref{fig:fig9} a comparison of a SM fit to the full set of 158\agev\ NA49 data~\cite{13,17} (Fig.~\ref{fig:fig9}~(a)), with a fit where $\bar{\mathrm{p}}$, $\bar{\Lambda}$ and $\bar{\Xi}$ were excluded (Fig.~\ref{fig:fig9}~(b)). Indeed, in accordance with the corresponding UrQMD exercise shown in Fig.~\ref{fig:fig4}, the resulting SM parameters move up in the temperature $T$ by about 25~MeV,
in the latter case, along with an improved chisquare.
However, this large shift is accompanied by a significant decrease of $\gamma_S$, which makes the 
conclusion of a major role of afterburning in antibaryon absorption questionable, because such a shift is
not observed in the simulated data, as has been mentioned. In order to put our conclusion on
a firmer ground, we have repeated the fit by fixing $\gamma_S$ at a pre-set value
of $0.85$ (see e.g.\ Refs.~\cite{2,11}) in both cases. The results are shown in Fig.~\ref{fig:fig10} (top and bottom panels). The shift in temperature when excluding antibaryons (except $\bar{\Omega}$) is smaller than in the fits shown in Fig.~\ref{fig:fig9}, but still significant.

\begin{figure}
\includegraphics[width=0.39 \textwidth]{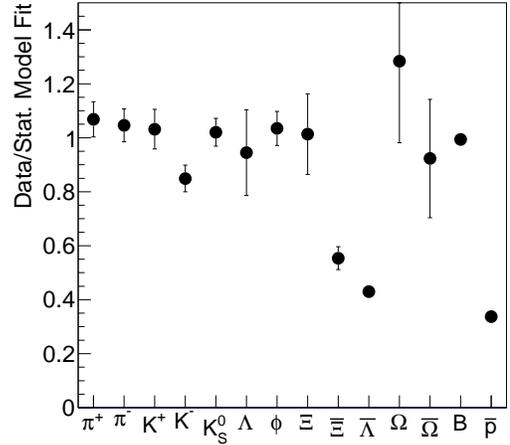}
\caption{Hadron multiplicity ratios data/SM implied by the statistical model fit
to the data set excluding the three antibaryons, shown in Fig.~\ref{fig:fig9}~(b).}
\label{fig:fig11}
\end{figure}

We show in Fig.~\ref{fig:fig11} that the differences between the data and the
statistical model fit, shown in Fig.~\ref{fig:fig10}~(b), exhibit
a striking resemblance to the corresponding ``hadron survival'' fractions, obtained from the 
UrQMD study illustrated in Fig.~\ref{fig:fig3}. Selective suppression of $\bar{\mathrm{p}}$, 
$\bar{\Lambda}$ and $\bar{\Xi}$ thus appears to be a property of the data.

A further indication of selective antibaryon absorption can be found in STAR Au+Au minimum 
bias data at $7.7~\gev$~\cite{16}, and in NA49 minimum bias 
data at $17.3~\gev$~\cite{17,18}. The kaon, $\Lambda$ and $\Xi$ multiplicities per participant 
nucleon increase smoothly toward central collisions, 
a property that can be well explained by 
the decreasing relative corona contribution~\cite{becacc,aichelin,bleicher}.
On the contrary, the $\bar{\mathrm{p}}$, $\bar{\Lambda}$ and $\bar{\Xi}$ yields per participant stay
constant with centrality in these collisions. One might thus infer on a further centrality dependent 
effect that works selectively in the latter cases. We recall the observation made in 
Fig.~\ref{fig:fig3} that the antibaryon absorption increases with centrality. This could 
counterbalance the yield increase per participant nucleon, exhibited by the baryons.

At the much higher top RHIC, and LHC energies, the selective suppression of the antibaryon yields (except 
for the $\Omega$ hyperon) in the hadron/resonance cascade stage, as implied in our results at SPS energies,
should give way to a suppression symmetric in baryons and antibaryons (as was already shown in Ref.~\cite{12}), 
due to the approximate particle-antiparticle symmetry prevailing here. 
Measurements of proton and antiproton midrapidity yields at LHC seem to confirm this \cite{Antinori:2011us}.
It is a likely possibility that the exclusion of those baryons from statistical model fits at
higher energy will also lead to an improvement of the fit quality and to a slight increase of 
the estimated chemical freeze-out temperature values. 

\section{Conclusions}

We have systematically investigated the effect of the hadron-resonance expansion phase that follows hadrochemical freeze-out, employing the hybrid version of the UrQMD transport model, which contains a transition process (from an intermediary hydrodynamic evolution), as described by the Cooper-Frye mechanism. We consider the hadron multiplicities obtained with, and without the cascade phase, analyzing them with the statistical model. We find that the cascade expansion as modeled in the microscopic dynamics of UrQMD does indeed modify the hadronic yield distribution, thus significantly affecting the position of the freeze-out curve obtained from the SM.
A downward shift occurs in the $(T,\mub)$ plane that, at first sight, appears to bridge the gap between the position of the lattice hadronization curve, and the well known statistical model freeze-out curve, at least at the investigated SPS energies. However, a closer analysis of the SM results shows that the observed shift is universally caused by a selective suppression of the antibaryons (except $\Omega$) during the cascade expansion stage. The other hadron yields stay very closely constant. What we observe is, thus, not a cooling of the hadron distribution but a distortion, and in fact it is accompanied by a systematic decrease of SM fit quality. If one omits these antibaryons from the SM fits most of the downward shift effect disappears.
This observation, which suggests a freeze-out curve staying close to the lattice estimate~\cite{5} of the parton-hadron coexistence line in the intermediate \mub\ domain, will be further checked 
as high statistics data from the STAR runs during the RHIC beam energy scan (BES) program become available.

An indication of analogous behavior is observed in SM analysis of the corresponding SPS data,
which needs to be confirmed by more detailed analyses, including a study of the centrality dependence
of the chemical freeze-out temperature at SPS and the use of the core-corona model to fit the data.
Also, the observed flatness of the centrality dependence of antibaryon yields per participant, 
in minimum bias collisions at this energy, may point to the presence of such a selective suppression.
This effect was underestimated in our
  previous publication~\cite{previous}, leading to a premature conclusion concerning
  the existence of an alternate QCD phase.

In summary these observations indicate that the hadronic freeze-out curve
shown in the QCD phase diagram of Fig.~\ref{fig:fig1} requires revision, at least, in the interval 
of SPS energies, at baryochemical potential from about 250 to 450~MeV, where the 
``empirical freeze-out curve'' from statistical model analysis begins to disentangle 
from the parton-hadron coexistence line proposed by
recent lattice QCD extrapolations to finite baryochemical potential.


This work was supported by the Deutsche Forschungsgemeinschaft (DFG), by 
the Hessian LOEWE initiative through HIC for FAIR and by the Istituto 
Nazionale di Fisica Nucleare (INFN). We are also grateful to the Center 
for Scientific Computing (CSC) at Frankfurt and to the INFN Sezione di
Firenze for providing the computing resources. T.~Schuster is grateful 
for support from the Helmholtz Research School on Quark Matter Studies. 
M.~Mitrovski acknowledges the support by the Brookhaven Science Associates, LLC under Contract No.\ DE-AC02-98CH1-8886 with the U.S.\ Department of Energy.
Moreover, this work was supported by GSI, BMBF and DESY.

\end{document}